# Manipulation of room-temperature Valley-Coherent Exciton-Polaritons in atomically thin crystals by real and artificial magnetic fields

Christoph Rupprecht[1], Evgeny Sedov[2,3,4], Martin Klaas[1], Heiko Knopf[5,6], Mark Blei[7], Nils Lundt[1], Sefaattin Tongay[7], Takashi Taniguchi[8], Kenji Watanabe[8], Ulrike Schulz[5], Alexey Kavokin[2,3,9], Falk Eilenberger[5,6,10], Sven Höfling[1,11], Christian Schneider[1]

*[1]Technische Physik and Wilhelm-Conrad-Röntgen-Research Center for Complex Material Systems, Universität Würzburg, D-97074 Würzburg, Am Hubland, Germany and Würzburg-Dresden Cluster of Excellence ct.qmat*

*[2]Westlake University, School of Science, 18 Shilongshan Road, Hangzhou 310024, Zhejiang Province, China*

*[3]Westlake Institute for Advanced Study, Institute of Natural Sciences, 18 Shilongshan Road, Hangzhou 310024, Zhejiang Province, China.*

*[4]Vladimir State University named after A.G. and N.G. Stoletovs, Gorky str. 87, 600000, Vladimir, Russia*

*[5]Fraunhofer Institute for Applied Optics and Precision Engineering IOF, Center of Excellence in Photonics, Albert-Einstein-Straße 7, 07745 Jena, Germany*

*[6]Institute of Applied Physics, Abbe Center of Photonics, Friedrich Schiller University, Albert-Einstein-Straße 15, 07745 Jena, Germany*

*[7]School for Engineering of Matter, Transport, and Energy, Arizona State University, Tempe, Arizona 85287, USA*

*[8]National Institute for Materials Science, Tsukuba, Ibaraki 305-0044, Japan*

*[9]Saint-Petersburg State University, Spin Optics Laboratory, Ul'anovskaya 1, Peterhof, St. Petersburg 198504, Russia*

*[10]Max Planck School of Photonics, Germany*

*[11]SUPA, School of Physics and Astronomy, University of St. Andrews, St. Andrews KY 16 9SS, United Kingdom*

**Strong spin-orbit coupling and inversion symmetry breaking in transition metal dichalcogenide monolayers yield the intriguing effects of valley-dependent optical selection rules. As such, it is possible to substantially polarize valley excitons with chiral light and furthermore create coherent superpositions of K and K' polarized states. Yet, at ambient conditions dephasing usually becomes too dominant, and valley coherence typically is not observable. Here, we demonstrate that valley**

**coherence is, however, clearly observable for a single monolayer of $WSe_2$, if it is strongly coupled to the optical mode of a high quality factor microcavity. The azimuthal vector, representing the phase of the valley coherent superposition, can be directly manipulated by applying magnetic fields, and furthermore, it sensibly reacts to the polarization anisotropy of the cavity which represents an artificial magnetic field. Our results are in qualitative and quantitative agreement with our model based on pseudospin rate equations, accounting for both effects of real and pseudo-magnetic fields.**

Atomically thin transition metal dichalcogenide crystals (TMDCs) belong to an emergent class of materials relevant to studies in fundamental- as well as application-oriented light matter interaction. They combine huge oscillator strength (optical activity) and giant exciton binding energies, making them a particularly interesting platform for cavity quantum electrodynamics. To date, the regime of strong light matter coupling in optical microcavities has been convincingly demonstrated with single monolayers[1], even at room temperature[2-4]. Moreover, TMDCs yield a plethora of peculiar symmetry-related effects. As a consequence of strong spin-orbit coupling and the broken inversion symmetry of the crystal, exciton spin orientations are inverted at opposite K points at the corners of the hexagonal Brillouin zone[5,6] . Hence , the K and K' valleys can be selectively addressed by $\sigma_+$ and $\sigma_-$ circularly polarized light[7-9]. This effect is termed excitonic spin-valley locking. The selective polarization of the valleys has been demonstrated by various resonant- and non-resonant optical techniques[10–13]. Interestingly, it has been shown that spin-relaxation can be suppressed in the strong light-matter coupling regime[13–18] since the relaxation dynamics is enhanced and exchange interactions are weakened.

The application of external magnetic fields lifts the energetic degeneracy of $\sigma_+$ and $\sigma_-$ polarized valley excitons (and trions) via the Zeeman-effect[19–23]. It has recently been shown, that the Zeeman-splitting of valley-excitons also persists in the strong coupling regime as a consequence of light-matter hybridization[24].

Here, we study the behavior of (exciton)-polaritons formed by strongly coupling a $WSe_2$ monolayer exciton to a high-Q photonic mode of a monolithic microcavity. When applying high magnetic fields up to 9 T, we observe that the characteristic valley Zeeman splitting of the K and K' valley excitons is preserved and transferred to the polaritons at room temperature, with an excitonic g-factor of 4. Upon excitation with linearly polarized light, coherent superpositions of valley polaritons can be created. First, we demonstrate, that the position of the polariton's polarization vector follows the polarization vector in the absence of a magnetic field. The lifting of the valley degeneracy by a magnetic field leads to a precession of the polaritons' polarization vector, and we demonstrate a rotation of the polarization vector by almost 90 °, during their relaxation towards the ground state. The angle-dependent linear polarization splitting of the microcavity also lifts the degeneracy of the polarization modes. It can thus be understood as an artificial magnetic field, which again yields a significant rotation of the polarization angle, by up to 280°. We have modeled the effect of both types of magnetic field on valley coherence in the monolayer with a system of linear rate equations for polariton polarization states. A good qualitative agreement between modelling and experiment has been obtained.

## Sample Structure and characterization

The studied sample structure is schematically depicted in Fig. 1a. The microcavity is built by transferring a $WSe_2$ monolayer with a dry-gel method[25] onto an $SiO_2/TiO_2$ bottom distributed Bragg reflector (DBR), grown onto a quartz glass substrate by physical vapor deposition. The bottom DBR consists of 10 pairs and has a stop band center at 1.653 eV. The monolayer was capped with a mechanically exfoliated flake of hexagonal boron nitride (h-BN, about 10 nm thick), to protect the monolayer from the subsequent overgrowth processing conditions. The top DBR (8 pairs) was grown by plasma-assisted evaporation (PAE) at mild processing temperatures of 80° C [26]. The complete, but empty microcavity exhibits a clear, parabolic dispersion relation with a ground mode at 1.658 eV as shown in the in-plane momentum-resolved reflectivity measurement (Fourier plane imaging method) in Fig 1c. The line spectrum at zero in-plane momentum (Fig. 1d) reveals a resonance linewidth $\Delta E$ of $(1.860 \pm 0.054)$ meV, equivalent to a Q-factor of $892 \pm 26$. Uncertainties were extracted from the fitting parameter of the Lorentzian. Fig. 1b shows a microscopy image of the final sample with the monolayer of $WSe_2$ and h-BN capping adjacent to bulk remnants taken before the growth of the top DBR.

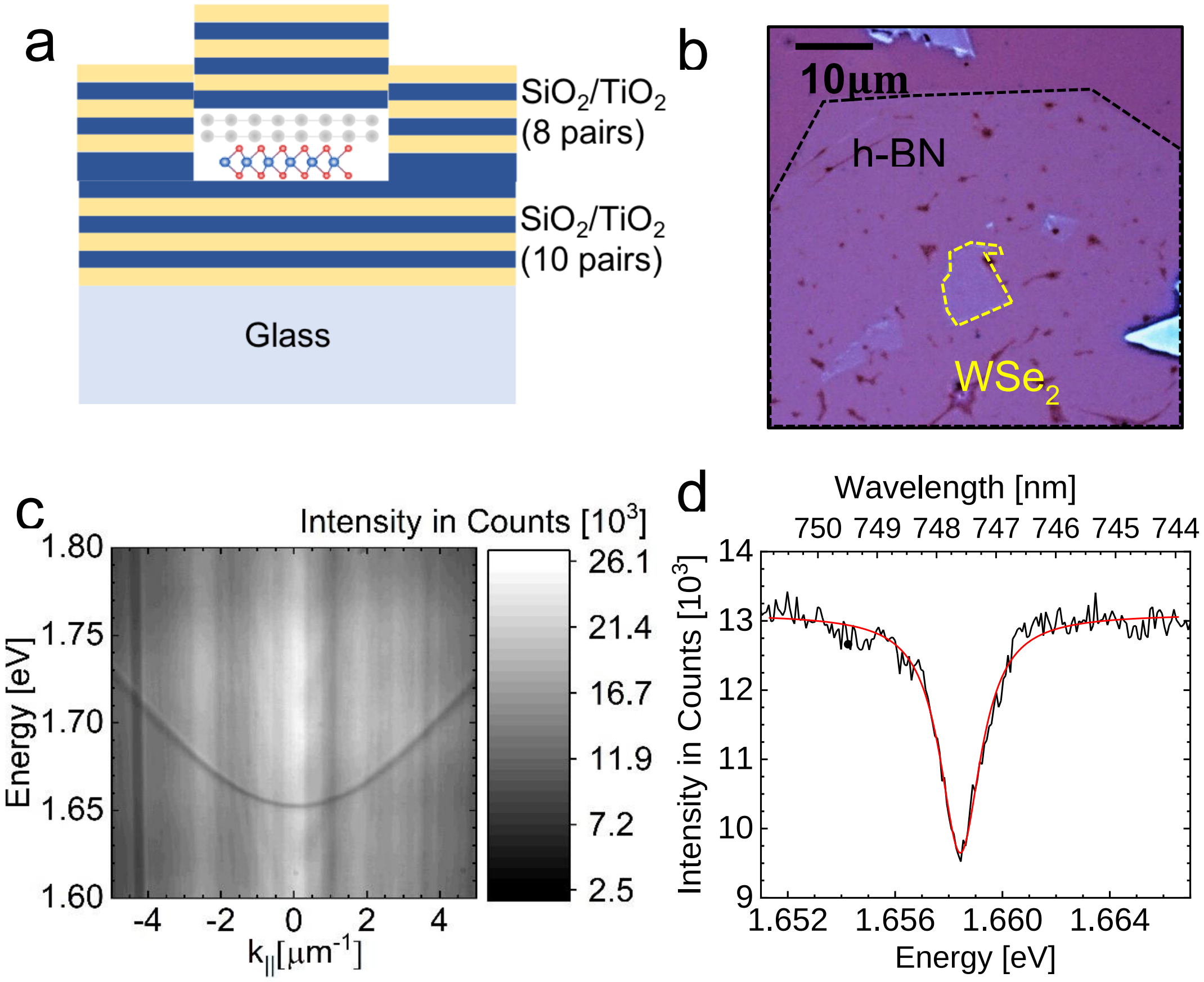

*Fig. 1 | The sample structure* a) Schematic of the fully grown microcavity structure with embedded $WSe_2$ monolayer capped by h-BN b) A microscope image of the capped $WSe_2$ flake c) The white light reflection spectrum showing the empty cavity resonance in momentum space d) The line spectrum of the empty cavity resonance at $k_{||} = 0\ \mu m^{-1}$ fitted by a Lorentzian revealing a $Q = 892 \pm 26$.

**TMDC based exciton-polaritons**

We study the luminescence properties of our $WSe_2$-cavity device via in-plane momentum-resolved photoluminescence. Figure 2a shows a photoluminescence (PL) spectrum of the completed microcavity sample, where we plot the emission energy as a function of the in-plane momentum. Each line spectrum (with constant $k_{||}$) was normalized to its maximum to better recognize the polariton emission at high $k_{||}$ more clearly. The empty cavity resonance (lower DBR/h-BN/upper DBR; stemming from areas outside the monolayer region) is visible at ~ 1.63 eV at $k_{||}$=0 $\mu m^{-1}$ (green solid line). Another empty cavity resonance (lower DBR/upper DBR) is weakly visible at ~ 1.655 eV at $k_{||}$=0 $\mu\text{m}^{-1}$ (see also Fig.1c). The extracted peak positions are plotted in Fig. 2b as function of $k_{||}$, following a well-pronounced parabolic dispersion. Accounting for the refractive index of the monolayer, the bare cavity resonance shifts to 1.610 eV ($k_{||}$=0 $\mu m^{-1}$) at the monolayer position (green dashed parabola in Fig 2a). Polaritonic modes arise on the red side of this mode with a significantly altered dispersion relation. This is best captured from the extracted peak position in Fig. 2b, which clearly indicate the change in dispersion slope towards large momenta. The energy spectrum of our device can be fitted with a standard coupled oscillator model[26],yielding a mode coupling (Rabi-Splitting) of $V = (70.7 \pm 3.7)$ meV. As input parameters, we use the properties of the uncoupled exciton, the cavity resonance, as well as the measured lower polariton branch, making the coupling strength the only free fitting parameter. The Hopfield coefficient determines the exciton fraction in the polariton state and its value is $|X|^2$= $0.125 \pm 0.015$ at $k_{||}$=0 $\mu m^{-1}$.

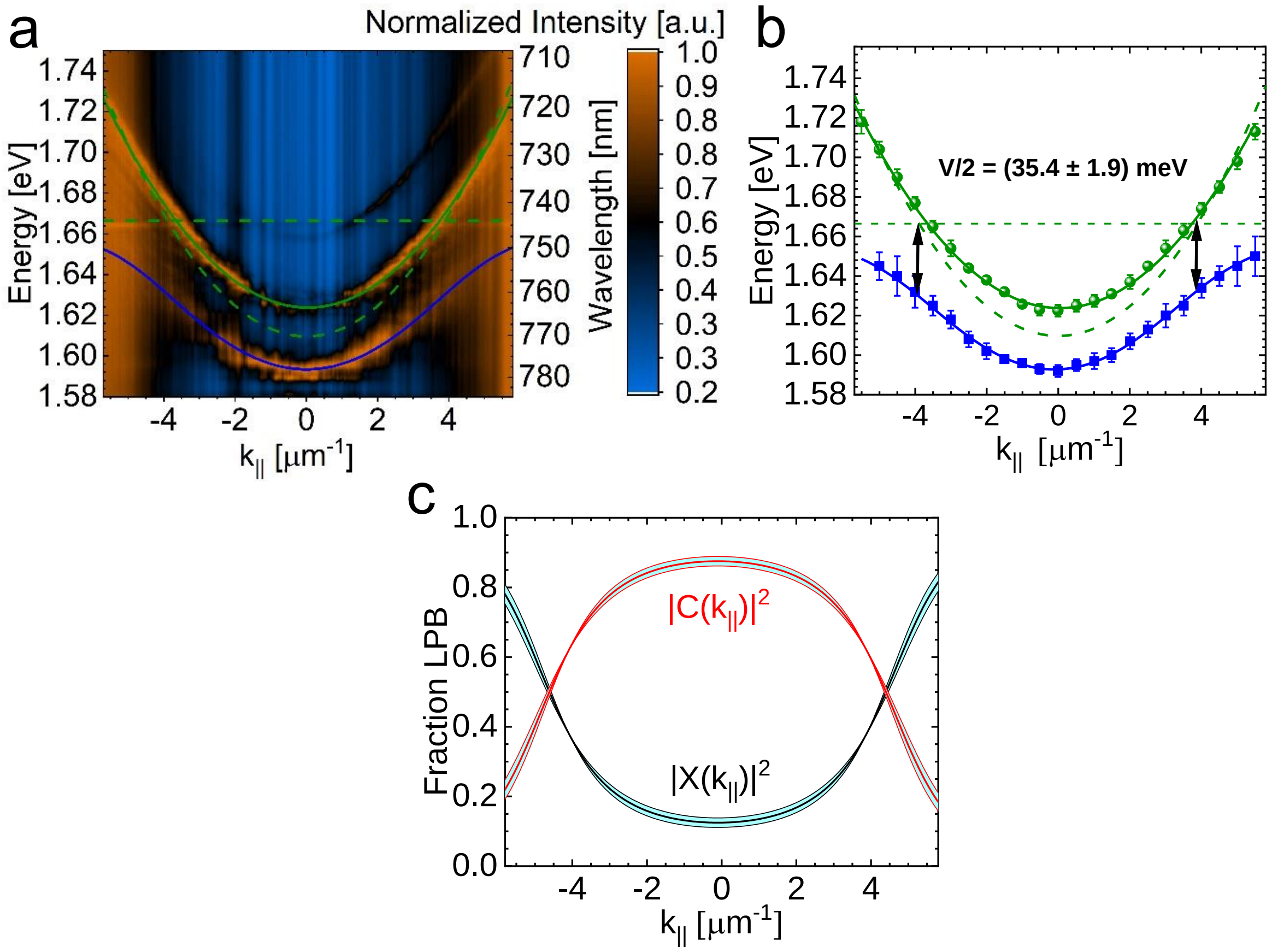

*Fig. 2* **/ Polariton dispersion and Hopfield coefficients** a) The normalized polariton dispersion in momentum space. The blue lines corresponds to a coupled oscillator fit with coupling constant $V = (70.7 \pm 3.7)$ meV. The green dashed lines show the exciton and the cavity resonance dispersion. The solid green parabola corresponds to resonance of a cavity with h-BN but without WSe2 b) The extracted peak positions with uncertainties from a) and the coupled oscillator fit with the same colors as in a and the upper polariton branch depicted in red c) Calculated Hopfield coefficients in momentum space with coincidence interval for the photonic (excitonic) portion $|C(k_{\|})|^2$ ($|X(k_{\|})|^2$) of the polariton. Uncertainties are extracted from the fit parameters of the coupled oscillator model.

## Polariton valley Zeeman splitting

As we have reported recently[24], the polaritons reflect the excitonic properties in that they exhibit a notable sensitivity to an external magnetic field. Thus, we investigate our completed microcavity sample in the presence of an applied magnetic field up to 9 T in the Faraday configuration. We conducted Stokes measurements by analyzing the polarization of the emission via rotating a quarter-wave plate and a linear polarizer. The PL-spectrum at $k_{\parallel} = (1.54 \pm 0.05)\ \mu m^{-1}$ and at a fixed magnetic field $B$ was fitted for each angle $\theta$ of the quarter wave plate by a Lorentzian and the peak positions $E(\theta)$ were extracted. These were then fitted with the equation

$$E(\theta) = C + D * \theta + \frac{\Delta \mathrm{E_{Pol}}\,(k_{\parallel})}{4} \cos(4\theta + \phi) + \frac{\Delta E_Z(B)}{2} \sin(2\theta) \tag{1}$$

where C is a constant. $\Delta E_Z(B)$ is the Zeeman splitting as a function of the external magnetic field $B$ and $\Delta \mathrm{E_{Pol}}(k_{\parallel})$ is the energetic separation of superimposed TE/TM polarizations of the structure as a function of the transverse angular momentum $k_{\parallel}$ of the emitted light. Note that the cavity TE/TM splitting is two order of magnitude larger than the excitonic longitudinal-transverse splitting, which is thus negligible[28,29]. $\phi$ takes into account the polaritons polarization rotation created by the TE/TM-splitting[30] and the linear term $D$ describes possible drifts during a measurement series. The emergence of this equation as well as typical observed measurements are discussed more in detail in the supplementary information.

Fig. 3c depicts the final results for the Zeeman-Splitting $\Delta E_Z(B)$. The Zeeman splitting increases linearly with the applied magnetic field up to a value of approximately 0.45 meV at $B = 9$ T. This is also reflected in the line spectra at 0 T and 7 T in Fig. 3a and b. These were extracted at $\theta = 70°$ and $\theta = 160°$ from the Stokes measurement, where only $\sigma^+$ and $\sigma^-$ polarized light reaches the spectrometer. $\Delta E_Z(B)$ can be described by [31]

$$\Delta E_z(B) = \frac{1}{2} g\mu_B B + \frac{1}{2}\sqrt{\delta_-^2 + V^2} - \frac{1}{2}\sqrt{\delta_+^2 + V^2} \tag{2}$$

with $\delta_{\pm} = E_C - E_X(B) \mp \frac{1}{2} g\mu_B B$. $g$ is the excitonic g-factor, $V$ is the coupling constant, $E_C$ and $E_X(B)$ are the cavity and photon energies, respectively. Since the diamagnetic shift of excitons in TMDC monolayers for magnetic fields <10 T is negligible[32], we can safely neglect the magnetic field dependence of $E_X(B)$ in our analysis. Fitting our data (red) yields an exciton g factor of $4.02 \pm 0.22$, which is in excellent agreement with results obtained on bare $WSe_2$ excitons in the existing literature[18-20]. Furthermore, the weakly coupled character of the mode, which we have attributed to the empty cavity resonance next to the monolayer (green solid line if fig 2a) is reflected by the absence of any observable splitting in Fig 3d. (Typical spectra are shown in the supplementary materials).

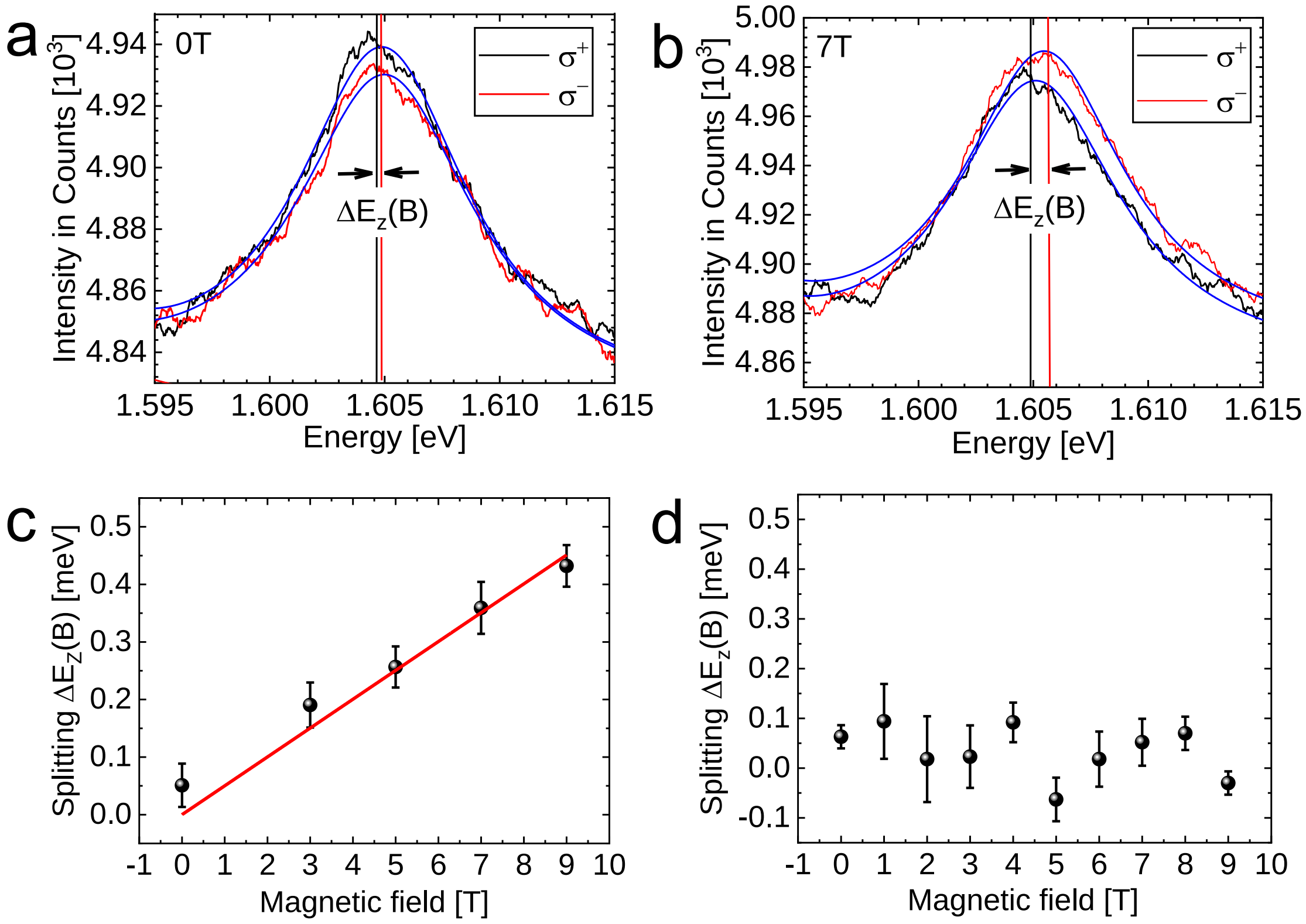

*Fig. 3 | Polariton Zeeman splitting*. a) and b) show spectra for the analyzed Zeeman peak at 0 T and 7T for 70° ($\sigma^+$) and 160° ($\sigma^-$) rotation of the quarter wave plate. Blue lines are Lorentzian fits to the data. c) Polariton Zeeman splitting $\Delta E_z(B)$ extracted from Stokes measurements at $k_\parallel = (-1.544 \pm 0.024)\ \mu m^{-1}$ for increasing magnetic fields. $g = (4.02 \pm 0.22)$ was extracted through a fit based on equation (2) (red). Uncertainties estimated from the Stokes fit. d) Absence of Zeeman Splitting of the photon mode shown in Fig. 2a) at $k_\parallel = (-1.44 \pm 0.05)\ \mu m^{-1}$ obtained at a position of the cavity without a TMDC.

## Room-temperature valley coherence

We now scrutinize the polarization properties of our structure in more detail. While significant degrees of valley polarization have been obtained in TMD-exciton-polariton structures both at cryogenic as well as room temperature, large degrees of valley coherence have only been recently found in non-resonantly as well as quasi resonantly excited structures at cryogenic temperatures[28,33]. First indications were also reported in a strongly coupled device at room temperature[34]. Here we have significantly extended this study and have demonstrated the room-temperature valley coherence can be widely manipulated by external fields.

Measurements at different magnetic fields were performed using a $\lambda/2$-wave plate and a following fixed horizontal linear polarizer in the detection path and the wave plate was rotated during the experiment, thus detecting the linear polarization component with respect to a fixed angle $\vartheta$. In Fig. 4a and 4b, we represent results at $B = 0$ T and at a transverse momentum of $(3.0 \pm 0.1)\ \mu m^{-1}$. The structure is driven by a linearly polarized cw laser at 532 nm. The degree of linear polarization (DOLP) is calculated as $(I_+ - I_-)/(I_+ + I_-)$, and reflects the degree of valley coherence retained in our sample. We make two important observations. First, in reasonable agreement with recent reports[34], the strong coupling conditions allow us to retain parts of the valley coherence, indicative by the DOLP of approx. 35 % (see also Fig 5b) in the luminescence of our device. Second, the azimuthal direction of the polarization vector follows the direction of the pump laser, which reflects that the phase between the K and K' exciton-polaritons is stable and externally tunable by the incoming pump (Fig 4).

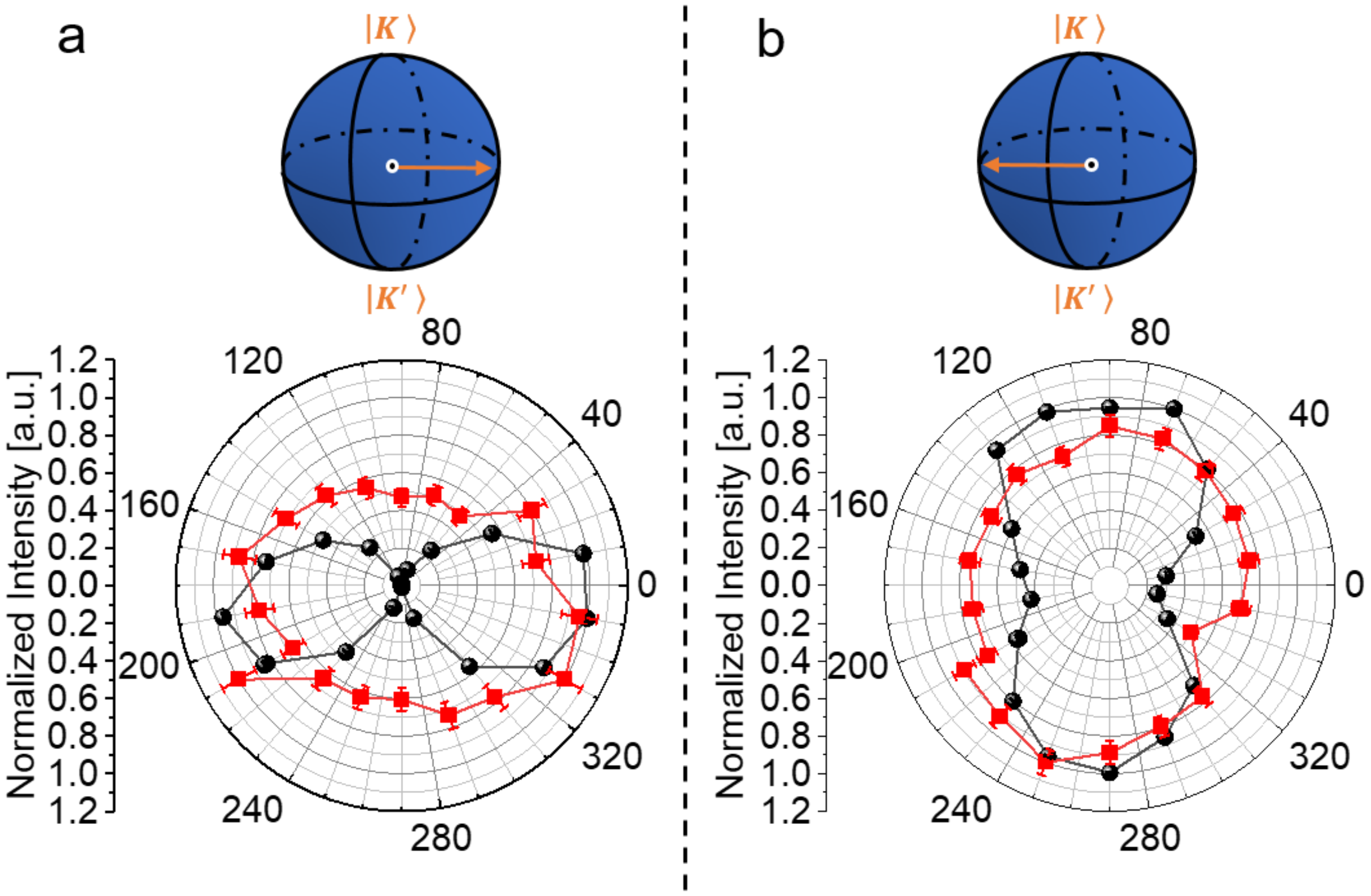

**Fig. 4 | Valley coherence** The laser polarization (black) and the polariton polarization (red) at $k_{||} = (3.0 \pm 0.1)\ \mu m^{-1}$ extracted from stokes measurements for different laser orientations. The uncertainties were estimated from the peak fits of the line spectra and the intensities normalized to the maximum of each Stokes series. The relative orientation of the polarization vector on the Bloch sphere is depicted.

We argue, that the phase between K and K' polaritons acquires a time-dependent term, if the energy degeneracy is lifted[35]. A straightforward way to break the degeneracy between K and K' polaritons has been introduced above, as the polaritonic valley Zeeman effect. We thus study both, the direction as well as the magnitude of the linear polarization emitted from our device at a fixed in-plane momentum of $(1.393 \pm 0.020)\ \mu m^{-1}$ as a function of the externally magnetic field applied field $B$. In Fig. 5a we plot the rotation of the polarization of the PL-light with respect to the laser polarization. We find up to 40° for $B = \pm 9\mathrm{T}$. The angles are extracted regarding to the horizontal axis in Fig. 4. This rotation is accompanied by a significant reduction of the valley coherence, which is reflected by the reduced DOLP plotted in Fig 5b.

In previous works, dedicated to the investigation of valley polarized exciton-polaritons in similar structures, it was argued that the energy splitting between TE and TM polarized cavity modes has a significant impact on the polarization properties[28,33]. Exciton-polaritons, being superposition states formed by cavity photons and excitons confined in the TMD monolayer, inherit properties of both their constituents, including the splitting in TE and TM modes. Due to the optical selection rules, $\sigma^+$ and $\sigma^-$ circularly polarized photons are able to selectively excite spin up and spun down excitons, respectively. The spin-valley locking being a characteristic feature of TMD monolayers leads to opposite spin orientations in opposite valleys. This allows one to associate the photon polarization with the exciton spin-valley degree of freedom. In the basis of circular polarizations (or spin up and spin down excitons), the Hamiltonian of the polariton $\mathbf{k}$-state takes the following form:

$$\widehat{H}_{\mathbf{k}} = \begin{pmatrix} \varepsilon(k) & \frac{1}{2}\Delta_{\mathrm{LT}}(k)e^{-2\mathrm{i}\theta} \\ \frac{1}{2}\Delta_{\mathrm{LT}}(k)e^{2\mathrm{i}\theta} & \varepsilon(k) \end{pmatrix},$$

where $\varepsilon(k)$ is the energy of the $\mathbf{k}$-state. The TE-TM splitting appears in this Hamiltonian as off-diagonal elements, where $\theta$ and $k$ are the angle and the wave number in the cavity plane characterizing the $\mathbf{k}$-state as $\mathbf{k} = (k\cos\theta, k\sin\theta)$. To ease notation, we omitted the subscript "$_{||}$" at $k$. $\Delta_{\mathrm{LT}}(k)$ is the magnitude of the TE-TM splitting, which in general case is $k$-dependent. The Hamiltonian (3) can be rewritten in a more elegant form:

$$\widehat{H}_{\mathbf{k}} = \varepsilon(k)\hat{\sigma}_0 + \mathbf{\Omega}_{\mathrm{C}} \cdot \hat{\mathbf{S}}, \tag{4}$$

where we introduced the three-dimensional polariton pseudospin operator $\hat{\mathbf{S}} = \frac{1}{2}\hat{\boldsymbol{\sigma}}$, with $\hat{\boldsymbol{\sigma}} = (\hat{\sigma}_x, \hat{\sigma}_y, \hat{\sigma}_z)$ being the vector of Pauli matrices. $\hat{\sigma}_0$ is the $2 \times 2$ identity matrix. In this form, the Hamiltonian $\widehat{H}_{\mathbf{k}}$ describes the precession of the polariton pseudospin around the effective in-plane magnetic field $\mathbf{\Omega}_{\mathrm{C}} = [\Delta_{\mathrm{LT}}(k)\cos 2\theta, \Delta_{\mathrm{LT}}(k)\sin 2\theta, 0]$ induced by the TE-TM splitting.

In our experimental configuration, we expect this effect to yield a similar rotation of the valley coherence as a real magnetic field. To make a quantitative analysis feasible, we first extract the linear polarization splitting of the polariton states as a function of the in-plane wave-vector, and plot it in Fig. 5c. To determine this splitting $\Delta E_s$, for every angle $\theta$ of the wave plate the energy position was identified by fitting a Lorentzian and these positions were adapted with

$$E(\theta) = C + D * \theta + \frac{\Delta E_s}{2}\cos(4\theta + \phi_1) \tag{5}$$

where C and D are constants and $\phi_1$ accounts again for a rotation of the linear polarization. This equation is discussed more in detail in the supplementary material as well as typical observed spectra are shown there. It is worth noting, that the magnitude of the effect strongly depends on the cavity TE-TM splitting and therefore on the choice of materials in the DBR mirrors, as well as on the mismatch of the cavity length from the Bragg condition. The strong splitting, which becomes as large as 1.5 meV at $k_{||}$-vectors of only 2 μm$^{-1}$, again yields a significant modification of the orientation of the linear polarization emitted from the polaritonic states. Indeed, the latter experience a significant rotation during their relaxation by an overall magnitude of approx. 280 °, which is more than five times stronger than the effect of the real magnetic field at 9T.

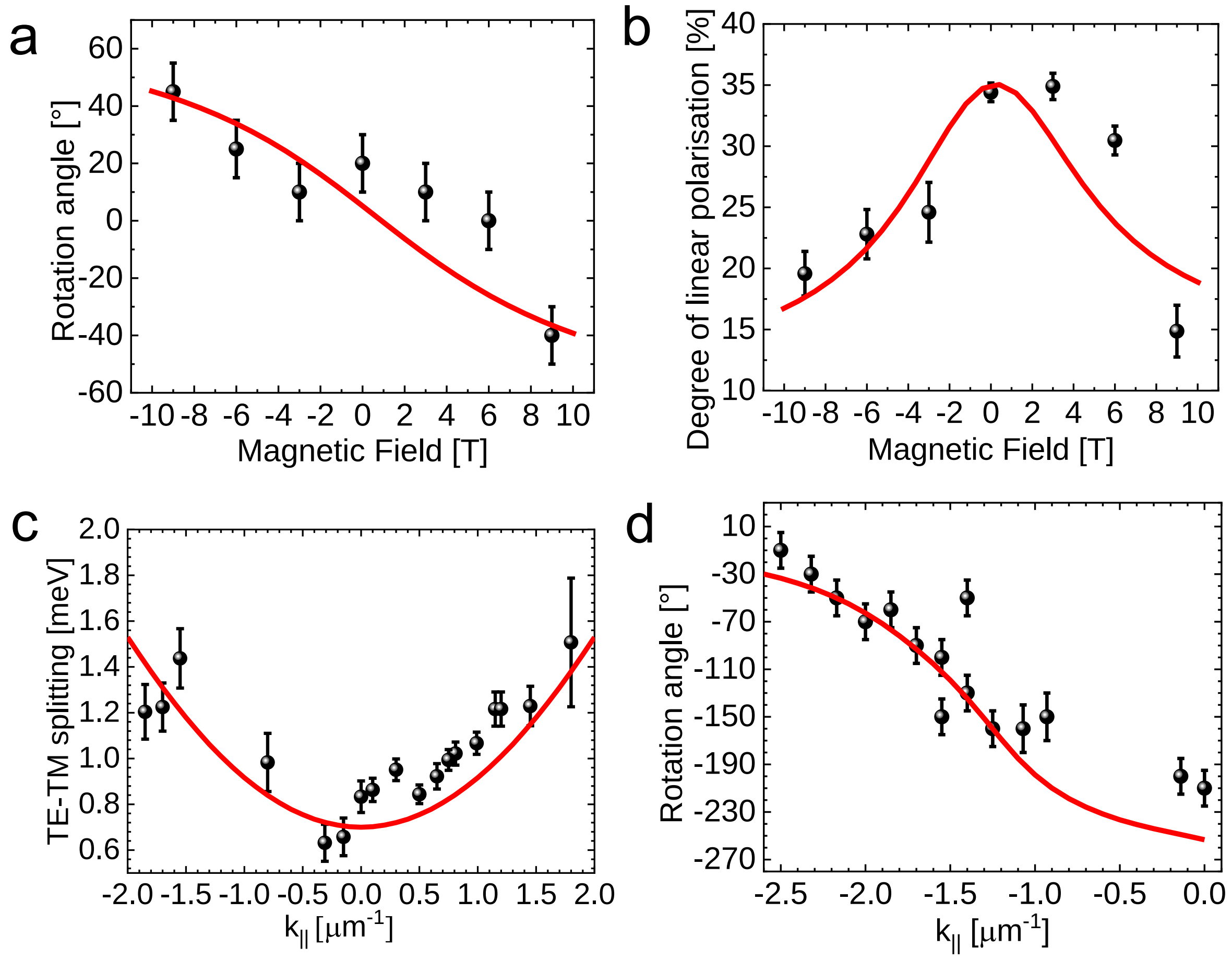

**Fig. 5 | Manipulation of valley coherence** a) Rotation angle of the linear polarization with respect to zero magnetic field for the polariton state at $k_{||} = (-1.393 \pm 0.020)\ \mu m^{-1}$ extracted from Stokes measurements. Uncertainties estimated from peak fits of the line spectra. b) The corresponding degree of linear polarization for the state, which rotation is shown in a). c) The linear polarization splitting at 0T extracted from Stokes measurements and parabolic fitted. Uncertainties extracted from the Stokes fit. d) The rotation angle related to horizontal from Fig.4a) at 0T for different $k_{||}$. The red curves in each panel show simulation data. The laser polarization is the same as in Fig. 4a).

Figure 5 also illustrates the manipulation of the valley coherence in the coupled monolayer-cavity structure reproduced numerically with a dynamical pseudospin model (red curves). The latter accounts for polarization-resolved pumping of polariton states, exciton and polariton pseudospin relaxation and quasimomentum-dependent lifetimes, as well as the effect of the effective magnetic field caused by the TE-TM splitting of cavity photons and the external magnetic field. The details of the models are given in the Supplementary materials section of the manuscript. As in the experiment, the linear polarization plane rotates with the applied magnetic field accompanied by the decrease in the polarization degree. The external magnetic field causes redistribution of populations of polariton pseudospin states and guides the system from the valley coherence to the valley polarization state. At zero magnetic field ($B = 0$), the only mechanism of pseudospin is the splitting in linear polarizations which is strongly inhomogeneous over the polariton dispersion, see Figs. 5c. Figure 5d shows the simulated rotation of the linear polarization plane with relaxation of polaritons down their dispersion.

## Conclusion

Our paper conveys two important messages: In the regime of strong light-matter coupling, valley coherence superpositions of exciton-polaritons directly manifest at ambient conditions, and under non-resonant pumping conditions. The phase of the coherence can be influenced- and controlled externally. We demonstrate that real- as well as artificial magnetic fields have a severe impact on the valley coherence, and represent efficient tools for its manipulation. Our findings show, that the effect of the TE-TM splitting on the polarization rotation in our microcavity is significantly stronger than the effect of an external magnetic field with an amplitude as large as 9T. Furthermore, it is straightforward to enhance and maximize the effect, e.g. by utilizing microcavities with a strong asymmetry[36], yielding strongly extended TE-TM splitting, or by engineering the natural birefringence reversibly, as has been demonstrated recently by applying external stress to semiconductor microcavities[37], or by utilizing strongly anisotropic liquid crystal filled DBR devices[38]. In such structures, it will be possible to gain control over the valley –pseudospin, its coherence, and finally the topology of valley polaritons in an unprecedented manner, by taking advantage of the coherent light-matter coupling.

## Acknowledgement

The Würzburg group acknowledges support by the state of Bavaria. C.S. acknowledges support by the European Research Commission (Project unLiMIt-2D, Grant Agreement No. 679288). This work has been supported by the Fraunhofer-Gesellschaft zur Förderung der angewandten Forschung e.V. F.E gratefully acknowledge the financial support by the German Federal Ministry of Education and Research via the funding "2D Nanomaterialien für die Nanoskopie der Zukunft". Work of E.S. and A.K. was supported by foundation of Westlake University (Project No. 041020100118 ). E.S. acknowledges partial support from the Grant of the President of the Russian Federation for state support of young Russian scientists No. MK-2839.2019.2. A.K. acknowledges the Saint-Petersburg State University for the research grant ID 40847559. SK.W and T.T. acknowledge support from the Elemental Strategy Initiative conducted by the MEXT, Japan and the CREST (JPMJCR15F3), JST. S.T. acknowledges support by the NSF (DMR-1955668 and DMR-1838443). H.K. is supported via the Max Planck School of Photonics.